\documentclass[runningheads]{llncs}

\makeatletter

\renewcommand\subsubsection{\@startsection{subsubsection}{3}{\z@}
  {-18\p@ \@plus -4\p@ \@minus -4\p@}
  {8\p@ \@plus 4\p@ \@minus 4\p@}
  {\normalfont\normalsize\bfseries\boldmath}}
\makeatother

\makeatletter
\renewcommand\paragraph{\@startsection{paragraph}{4}{\z@}
  {-12\p@ \@plus -4\p@ \@minus -4\p@}
  {-0.5em \@plus -0.22em \@minus -0.1em}
  {\normalfont\normalsize\bfseries}}
\makeatother

\usepackage[T1]{fontenc}
\usepackage{graphicx,verbatim}
\usepackage{amssymb}
\usepackage{amsmath}
\begin{document}
\title{Protocol generalisation for brain tissue microstructure estimation via hypernetwork-controlled geometric deep learning}
\titlerunning{Protocol generalisation via hypernetwork-controlled geometric deep learning}
%

\author{Andrea Brigliadori$^1$, Leevi Kerkela$^1$, Hui Zhang$^1$}
\authorrunning{A. Brigliadori et al.}
\institute{$^1$Hawkes Institute and Department of Computer Science, University College London, London, United Kingdom \\
    \email{andrea.brigliadori.24@ucl.ac.uk}}
  
\maketitle              
\begin{abstract}
Brain tissue microstructure estimation with machine learning provides higher computational efficiency than conventional fitting. However, machine learning still presents important limitations that hamper its clinical utility. Specifically, current models typically lack generalisation across diffusion MRI acquisition protocols and require retraining whenever b-vectors or b-values change. Moreover, the recent machine learning methods that were developed to address protocol generalisation lack rotational equivariance. Particularly suitable for dMRI parameter estimation is a geometric deep learning model known as spherical convolutional neural network (SCNN), which guarantees rotational equivariance and b-vector generalisation. However, this architecture currently does not account for b-values.
Therefore, obtaining a model that combines protocol generalisation and rotational equivariance remains an open challenge.
In this paper, we directly address this issue by incorporating explicit b-value dependence into an SCNN architecture via a hypernetwork. 
This new approach is illustrated using NODDI as an example forward model for estimating brain tissue microstructure. To evaluate b-value generalisation, the original and newly proposed SCNN architectures are trained on synthetic data and tested on both synthetic and real data across different b-value pairs.
Results demonstrate that the proposed method achieves reduced RMSE and bias on synthetic data, as well as higher agreement with conventional NODDI fitting on real data, indicating improved robustness to unseen b-values and a reduced need for retraining. By combining generalisation across b-values with generalisation across b-vectors and rotational equivariance, the proposed framework enhances the applicability of deep learning to clinical diffusion MRI parameter estimation. Code is publicly available at \texttt{https://github.com/aerdnairo/arXiv\_generalisedSCNN}.


\end{abstract}
\section{Introduction}
Brain tissue microstructure refers to the organisation of brain tissue at the micron scale. This microstructure can be probed with an MRI technique known as diffusion-weighted MRI (dMRI) \cite{OriginaldMRIPaper_LeBihan}, which works by using diffusion gradients to amplify the sensitivity of MRI signals to the random motion of water molecules. Each of these gradients is characterised by a strength, or b-value, and a direction, or b-vector, the two main scanner settings defining a dMRI acquisition protocol. Given these settings and the dMRI images acquired under these conditions, one can estimate microstructural properties in each voxel. The estimates can then be visualised as quantitative maps over the brain, and anomalies in such maps are helpful for the early detection of microstructure alterations due to multiple sclerosis \cite{dMRI_in_MS}, stroke \cite{dMRI_utility}, and brain tumours \cite{dMRI_in_braintumour}.

Estimating microstructure properties from the signals measured in a voxel is an inverse problem of dMRI. This problem is conventionally approached by assuming a forward model that maps tissue properties and acquisition settings to signal predictions. The properties are then estimated by minimising the distance between predictions and measurements.
However, solving this minimisation problem typically requires an iterative approach and one such problem must be solved separately for each of the hundreds of thousands of voxels. As a consequence, it is computationally expensive, requiring many hours for a whole-brain estimation, thus limiting clinical feasibility \cite{Granziera2021Quantitative,NODDI4ClinicalResearch,Sam,NODDI}.

Machine learning (ML) offers an alternative paradigm in which the mapping from measurements to properties is learned from data. Once the mapping is found, it can be applied to any new voxel to obtain estimates in clinically practical times~\cite{Aliotta_EfficiencyofDL,Golkov_EfficiencyofDL}.  
However, early ML approaches produce mappings, also known as models, for specific acquisition protocols \cite{CommonML_5,CommonML_2,CommonML_1,CommonML_3,Sam}, thus requiring retraining whenever the protocol changes, which may hamper their clinical utility \cite{Professor_HowCanSCNNs,Granziera2021Quantitative,FromResearchToClinicalPractice,Sam}. 
Several more recent ML methods have been developed to address this limitation \cite{DIFFnet_bValGen,Zong_AttentionBasedQSpaceDL}, but they are typically not rotationally equivariant \cite{Professor_HowCanSCNNs,Kerkela}.

Rotational equivariance is key to guaranteeing that parameter estimates remain invariant to the pose of the brain, but introducing it a posteriori into a model is difficult. Spherical convolutional neural networks (SCNNs) ~\cite{Cohen_SCNNs,Esteves_SCNNs} naturally achieve rotational equivariance \cite{Cohen_SCNNs,Esteves_SCNNs,Kerkela,Sedlar} and were recently shown to generalise across different b-vector schemes~\cite{Professor_HowCanSCNNs}. These properties make SCNNs a natural starting point for improving the clinical utility of ML for dMRI parameter estimation. However, SCNNs still lack b-value generalisation.

In this paper, we propose what is, to the best of our knowledge, the first deep learning model to attain both rotational equivariance and protocol generalisation. We achieve this by combining an SCNN architecture with a hypernetwork~\cite{hypernetworks} that takes b-values as input and returns weights for the SCNN. This work extends our preliminary publication~\cite{Brigliadori2026}.

The goal of our experiments is to show that the hypernetwork-controlled SCNN (hSCNN) matches, for each choice of b-values, the performance of an SCNN trained specifically for that protocol. The proposed approach naturally works for any forward model, and is here illustrated with the example of NODDI because of its wide uptake in the community \cite{FWFAlzheimer,ODIVal3NDIVal1,NODDInicepicturesAndHistVal,ODIVal1}. We synthesise the training data for hSCNN using multiple b-value pairs. Then, for comparisons, we separately train multiple implementations of the baseline SCNN on data synthesised with either fixed or variable b-value pairs.
 Finally, we assess b-value generalisation by testing the hSCNN and the baseline SCNNs on synthetic and in-vivo data across different choices of b-values.  
 
Our results show that the bias and RMSE of hSCNN consistently match those of an SCNN trained for the same b-values used in the test set. Moreover, on in-vivo data, hSCNN produces parametric maps which closely agree with those derived from NODDI fitting. These findings demonstrate improved b-value generalisation and support the applicability of ML to clinical dMRI parameter estimation.

The rest of the paper is organised as follows: Section \ref{Sec:Theory} formally presents the inverse problem addressed in this paper, previous approaches to finding a solution and the method we propose. 
Section \ref{Sec:Methods} explains the specific model implementations, the data synthesis procedures and the evaluation steps.  
Results on synthetic and real data are then described in Section \ref{Sec:Results}. Finally, Section \ref{Sec:Conclusion} reports the discussion and conclusions. 

\section{Theory}\label{Sec:Theory}
This section formulates the inverse problem investigated in this work and describes how it is addressed using conventional fitting and machine learning. The limitations of current ML methods are then discussed, motivating the approach proposed in this work. 
\subsection{Inverse Problem Formulation}
  
Let $M$ dMRI measurements produce, for each voxel, signals $\mathbf{s}\in\mathbb{R}^M$. These depend on $P$ microstructural properties $\mathbf{x}\in\mathbb{R}^P$ and on the acquisition protocol $\mathbf{k}$, defined by b-values $\mathbf{b}=[b_1,\dots,b_M]^\top$ and b-vectors $\mathbf{G}=[\hat{\mathbf{g}}_1,\dots,\hat{\mathbf{g}}_M]\in\mathbb{R}^{3\times M}$. Given $\mathbf{s}$ and $\mathbf{k}$, the objective of computational modelling for dMRI is to find $\mathbf{x}$. Two main approaches exist to reach this objective, each proposing a different framework of the problem.

\subsection{Existing Solutions and Their Limitations}

\subsubsection{The Conventional Fitting Framework}
\begin{figure}[h!]
    \centering
    \textbf{The Conventional Fitting Framework} \\[0.5em]    
    \includegraphics[width=0.97\linewidth]{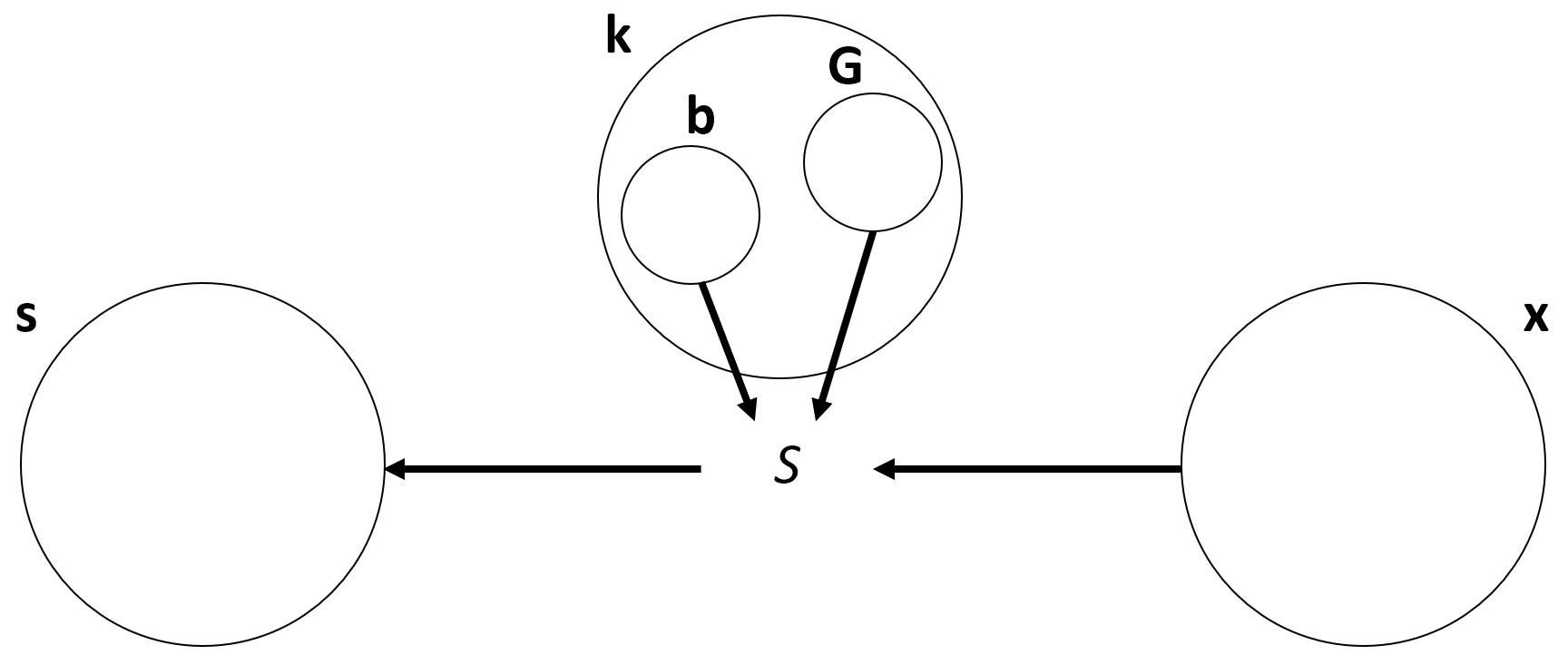}
    \caption{The conventional fitting framework assumes the
knowledge of a forward model S. This function takes as input $\mathbf{k}$, is parametrised by $\mathbf{x}$, and
returns predictions for $\mathbf{s}$.}
    \label{fig:ConventionalFitting}
\end{figure}

The traditional approach (Figure \ref{fig:ConventionalFitting}) assumes a forward model $S$ which takes as input $\mathbf{k}_j$, is parametrised by $\mathbf{x}$, and outputs a prediction of $s_j$, for a generic $j \in \{1, \dots, M\}$. The inverse problem is then restated as follows: Given $\mathbf{s}$, $\mathbf{k}$, and the forward model $S$, find $\mathbf{x}$. 
For each voxel, the solution $\hat{\mathbf{x}}$ is obtained by iteratively minimising the distance between predictions and measurements:
\begin{equation}
\hat{\mathbf{x}} = \arg \min_{\mathbf{x}}\sum_{j=1}^{M}{(s_j - S(\mathbf{k}_j; \mathbf{x}))^2}
\end{equation} 
The main limitation of this framework is the computational burden of voxel-wise iterative fitting, motivating ML approaches.

\subsubsection{The Machine Learning Framework}
\begin{figure}[h]
    \centering
    \textbf{The Machine Learning Framework for dMRI} \\[0.5em]    
    \includegraphics[width=0.97\linewidth]{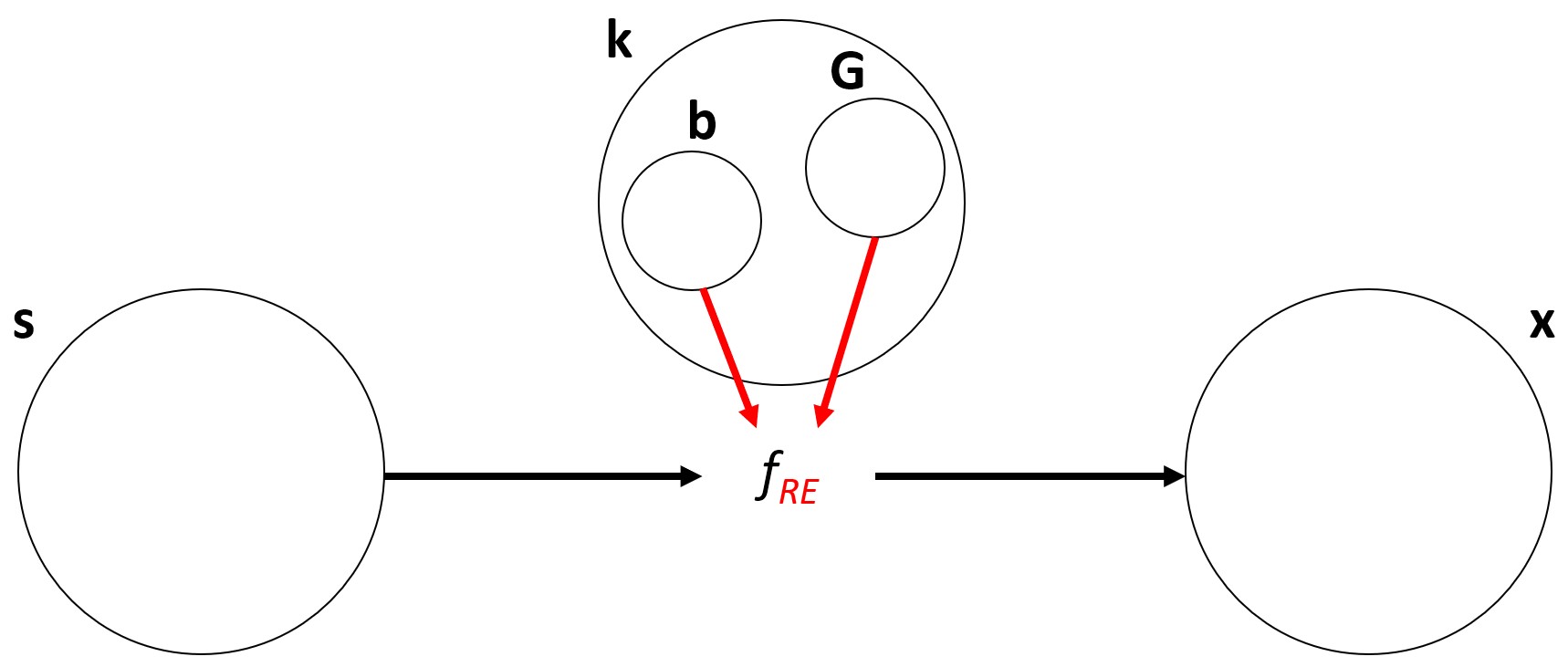}
    \caption{Machine learning aims to find the mapping between measurements $\mathbf{s}$ and tissue properties $\mathbf{x}$ from data. For dMRI applications, this mapping should also account for acquisition protocols and be rotationally equivariant.}
    \label{fig:IdealML}
\end{figure}

In the ML framework (Figure \ref{fig:IdealML}) the mapping between measurements and tissue properties is found from data. The function thus obtained is the ML model. To achieve clinical applicability, it is essential for the model to account for acquisition protocols and to be rotationally equivariant \cite{Professor_HowCanSCNNs,Granziera2021Quantitative,Sam}. Indeed, protocol generalisation allows the model to be used across choices of b-values and b-vectors, and rotational equivariance ensures that the estimation quality is maintained on unseen fibre directions \cite{Professor_HowCanSCNNs,Kerkela}.
Therefore, we represent the data used by the ML framework for dMRI as $N$ examples $\{(\mathbf{k}^{(i)}, \mathbf{s}^{(i)},\mathbf{x}^{(i)})\}_{i=1}^N$, where each example comprises an acquisition protocol, the corresponding voxel signals, and the underlying tissue properties. The framework further considers a rotationally equivariant (RE) function $f_{RE}$ parametrised by weights $\boldsymbol{\theta}$. The goal is to find $\hat{\boldsymbol{\theta}}$ such that: 
\begin{equation} \label{MLProblem}
\hat{\boldsymbol{\theta}}=\arg \min_{\boldsymbol{\theta}}\sum_{i=1}^{N}{\lVert f_{RE}(\mathbf{k}^{(i)}, \mathbf{s}^{(i)};\boldsymbol{\theta}) - \mathbf{x}^{(i)} \rVert^2}.
\end{equation}
This problem is solved via backpropagation \cite{backprop}, in a process referred to as training. Differently from the conventional fitting process, training can be done simply once, away from the clinic. Once the weights are found, the function can be applied to new voxels and provide immediate microstructure estimation. 
However, in practice, ML approaches for dMRI have often overlooked acquisition protocols and rotational equivariance, limiting the clinical utility of the proposed models.

\begin{figure}[h]
    \centering
    \textbf{Common Machine Learning Approaches} \\[0.5em]    
    \includegraphics[width=0.97\linewidth]{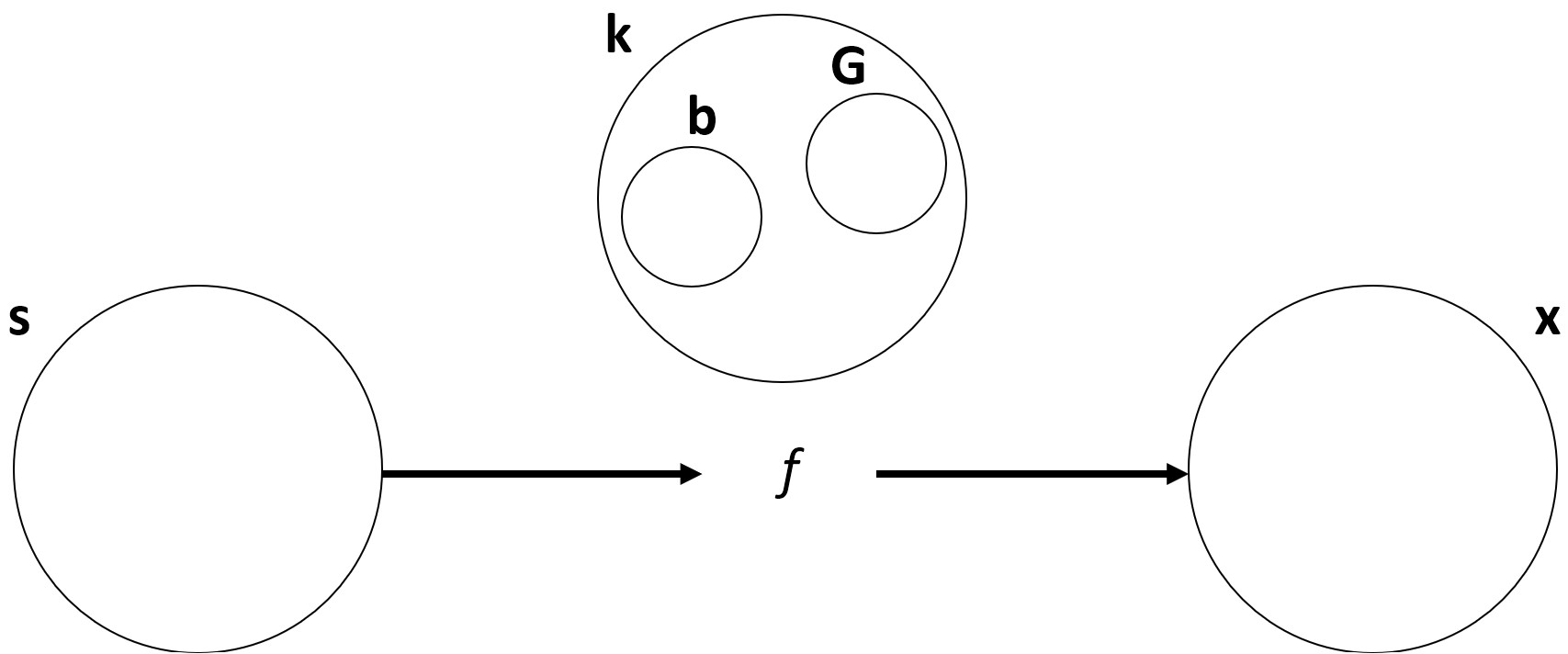}
    \caption{Often, machine learning is applied to dMRI parameter estimation overlooking the importance of acquisition protocols and rotational equivariance. }
    \label{fig:CommonML}
\end{figure}

\paragraph{Common Machine Learning Approaches} Often, ML approaches directly apply standard architectures such as multi-layer perceptrons (MLPs) to the signals, without accounting for the acquisition protocols \cite{CommonML_5,CommonML_2,CommonML_1,CommonML_3,Sam} (Figure \ref{fig:CommonML}). Such approaches consider $N$ examples $\{(\mathbf{s}^{(i)},\mathbf{x}^{(i)})\}_{i=1}^N$ of voxel signals paired with the corresponding properties, and a non-rotational equivariant $f$. Then, the following minimisation problem is solved:
\begin{equation}
\hat{\boldsymbol{\theta}}=\arg \min_{\boldsymbol{\theta}}\sum_{i=1}^{N}{\lVert f(\mathbf{s}^{(i)};\boldsymbol{\theta}) - \mathbf{x}^{(i)} \rVert^2}.
\end{equation}
As a consequence of this formulation, the weights $\hat{\boldsymbol{\theta}}$ are optimised for the training distributions of protocols and fibre directions, with no generalisation guarantees. Recently, ML models have been specifically designed to address this limitation.

\begin{figure}[h]
    \centering
    \textbf{Protocol-aware Machine Learning Approaches} \\[0.5em]    
    \includegraphics[width=0.97\linewidth]{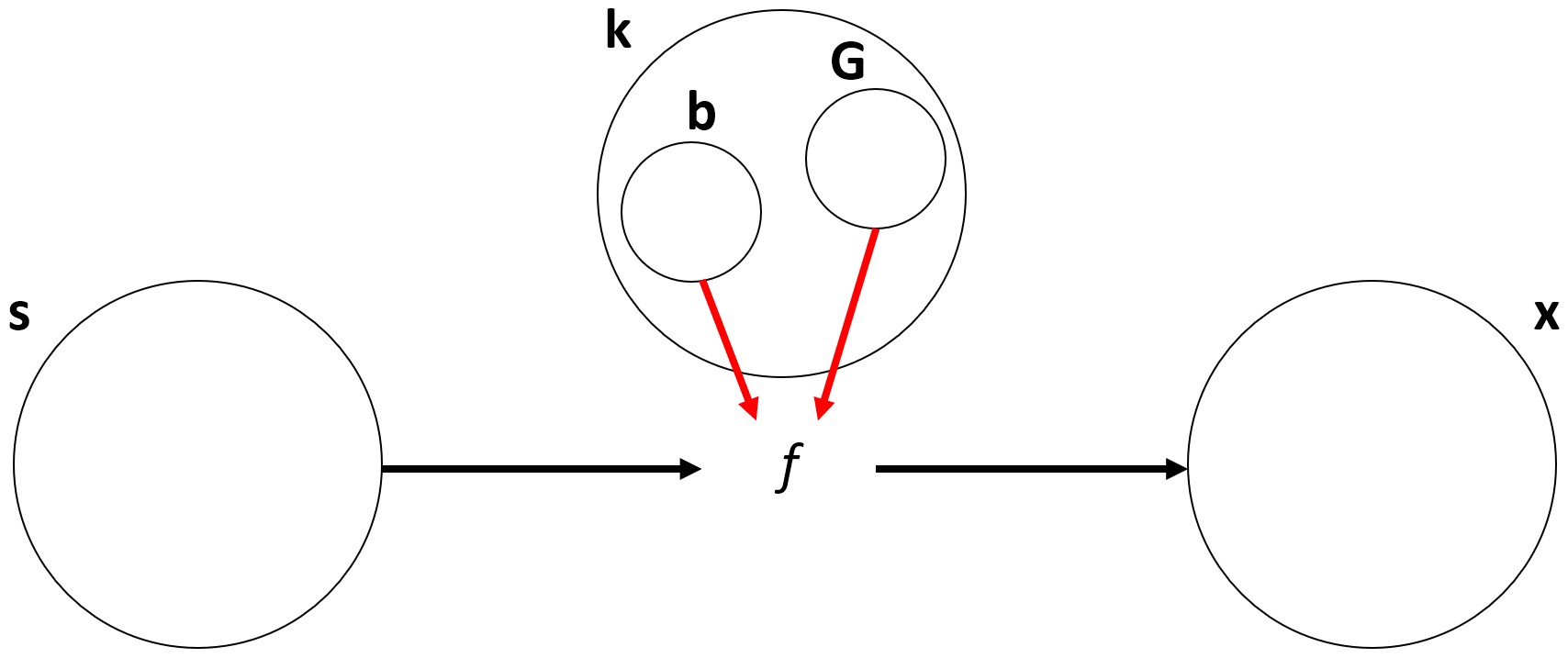}
    \caption{Recent machine learning models have been designed to generalise across b-values and b-vectors, but lack rotational equivariance.}
    \label{fig:ProtocolAwareML}
\end{figure}

\paragraph{Protocol-aware Machine Learning Models} An intuitive approach to account for b-values would be concatenating them to the input of simple neural networks such as MLPs. However, to account for both b-values and b-vectors, dedicated architectures have been proposed. In particular, DiffNet~\cite{DIFFnet_bValGen} represents diffusion measurements through a projected and quantised input matrix. More recently, aqDL~\cite{Zong_AttentionBasedQSpaceDL} employs Transformer encoders to map the measurements acquired with different protocols into a common feature space, using both b-values and b-vectors as input. According to these approaches, the minimisation problem is reframed as follows:
\begin{equation}
\hat{\boldsymbol{\theta}}=\arg \min_{\boldsymbol{\theta}}\sum_{i=1}^{N}{\lVert f(\mathbf{k}^{(i)}, \mathbf{s}^{(i)};\boldsymbol{\theta}) - \mathbf{x}^{(i)} \rVert^2}
\end{equation}
(Figure \ref{fig:ProtocolAwareML}), where the difference from Equation \ref{MLProblem} is that the proposed $f$ is not rotationally equivariant.
However, rotational equivariance is difficult to incorporate into a model a posteriori and is essential for maintaining estimation quality across unseen fibre orientations.

\paragraph{A Rotationally Equivariant Architecture: SCNNs} Spherical convolutional neural networks (SCNNs) \cite{Cohen_SCNNs,Esteves_SCNNs} are CNNs \cite{CNNs} that employ, at least in one layer, a spherical convolution. This operation allows SCNNs to achieve rotational equivariance, motivating our choice to focus on SCNNs as the state-of-the-art architecture for microstructure estimation, rather than the alternatives described above.

\begin{figure}[h]
    \centering
    \textbf{Rotational Equivariant and b-vector Generalisable Approaches} \\[0.5em]    
    \includegraphics[width=1\linewidth]{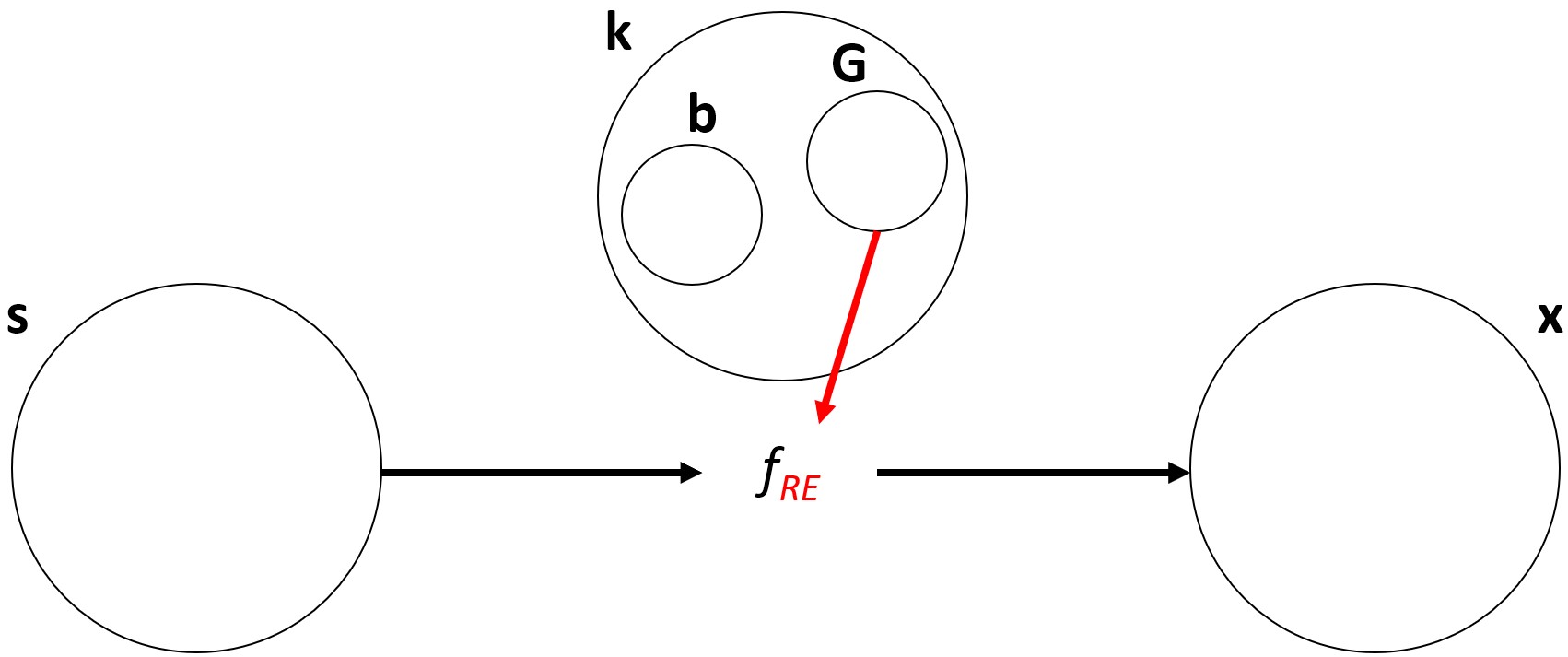}
    \caption{SCNNs naturally achieve rotational equivariance and b-vector generalisation.}
    \label{fig:SCNNs}
\end{figure}

In dMRI applications, SCNNs are applied voxel by voxel on a dataset of spherical harmonic (SH) coefficients computed from the signals \cite{Kerkela}. Since these coefficients are defined through spherical harmonics evaluated at the gradient directions, they provide a structured description of how the signal varies with b-vectors. Because of this, SCNNs are also robust to unseen gradient schemes \cite{Professor_HowCanSCNNs}, reducing the need for retraining when the b-vectors vary (Figure \ref{fig:SCNNs}).
However, SCNNs do not yet account for b-values. Consequently, in this case the minimisation problem can be formulated as follows: 
\begin{equation}
\hat{\boldsymbol{\theta}}=\arg \min_{\boldsymbol{\theta}}\sum_{i=1}^{N}{\lVert f_{RE}(\mathbf{b}^{(i)}, \mathbf{s}^{(i)};\boldsymbol{\theta}) - \mathbf{x}^{(i)} \rVert^2}.
\end{equation}
Despite the advantages presented by SCNNs, they are still limited by their need to be retrained whenever the diffusion weighting changes.

\subsection{Protocol Generalisation and Rotational Equivariance: The Proposed Approach}\label{Sec:Contribution} 
\begin{figure}
\centering
\includegraphics[width=1\linewidth]{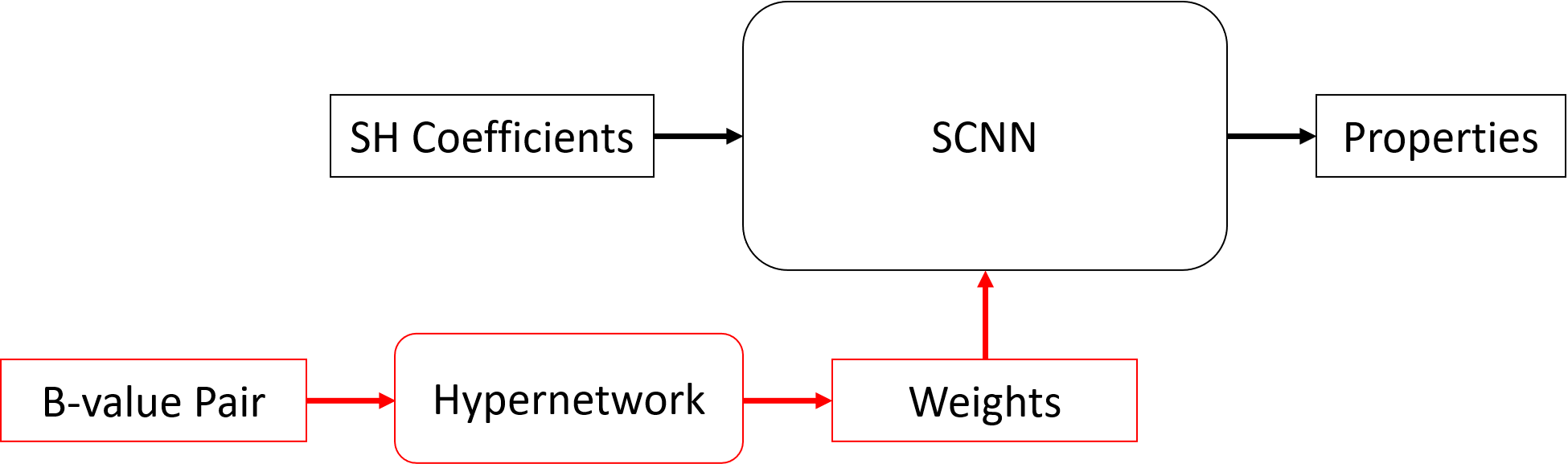}
    \caption{Diagram of the proposed model, combining b-vector generalisation, rotational equivariance, and b-value generalisation, with the latter achieved via a hypernetwork (red pathway).}
    \label{fig:MyhyperSCNN}
\end{figure}

Taken together, existing approaches provide either protocol generalisation or rotational equivariance, but not both simultaneously. To the best of our knowledge, the method proposed in this paper is the first that combines the two properties, instantiating the learning problem defined in Equation \ref{MLProblem}. This is achieved by leveraging the b-vector generalisation and rotational equivariance of SCNNs, while accounting for b-values via a hypernetwork \cite{hypernetworks} (Figure \ref{fig:MyhyperSCNN}).

A hypernetwork is a meta-learner which can be deployed on b-values to generate protocol-derived weights for a downstream ML model, effectively enabling it to adapt its estimates based on the b-values. In our case, this approach is particularly useful because it keeps the SCNN's required spherical harmonics input intact. Instead, simply concatenating scalar b-values into the input would violate the model's assumptions.

Remarkably, the proposed framework is neither tied to a specific forward model used to synthesise the data, nor to a specific architecture for the downstream network. In fact, the same approach could even be employed to introduce adaptability to further acquisition variables, such as SNR. The versatility of this approach facilitates its adoption across studies and clinical settings, positioning it as a compelling generalisation method for ML in diffusion MRI and other imaging modalities.

\section{Demonstration that the hypernetwork introduces b-value generalisation}\label{Sec:Methods}
In the following experiments, we focus solely on SCNN architectures as state-of-the-art microstructure estimators, due to their inherent rotational equivariance and b-vector generalisation. Our primary goal is to show that incorporating a hypernetwork (hSCNN) additionally confers b-value generalisation.

Ideally, an hSCNN trained across multiple b-value schemes should match the inference performance of SCNNs trained specifically for each protocol. To evaluate this, we compare an hSCNN trained on varied b-value pairs against both a baseline SCNN trained on the same multi-protocol data and two specialised SCNNs trained separately for each specific b-value pair.

Training datasets were synthesised using NODDI as an example forward model. We then evaluated b-value generalisation on both synthetic and in-vivo test sets featuring different b-value choices. The remainder of this section details the network implementations, data synthesis procedure, training hyperparameters, and evaluation steps.

\subsection{Model Architectures}
Following previous work~\cite{Kerkela}, we implemented an SCNN including six spherical convolution layers with channel sizes $[2,16,32,64,32,16,1]$ (input to output) and 78,387 parameters. Rotationally invariant scalar predictions were obtained by applying global pooling to the outputs of the first three convolution layers, followed by fully connected layers of sizes $[112,128,128,3]$ to predict the NODDI scalar parameters: NDI, ODI, and FWF.  

In the specific hSCNN architecture used in this paper, the hypernetwork was an MLP returning weights for the first spherical convolution layer. Using the hypernetwork only for the first layer was a deliberate choice to minimise complexity and prevent overparameterisation. The hypernetwork had input size 2, hidden size 8, and output size 1152 (corresponding to the target layer weights). The full hSCNN had a total of 80,715 parameters. 
\subsection{Data}
\subsubsection{Simulation of Synthetic Data}
All data was simulated using the NODDI MATLAB Toolbox~\cite{NODDIMATLABToolbox}, with parallel diffusivity fixed to $1.7 \times 10^{-3}\;\mathrm{mm^2/s}$ to match the standard NODDI implementation. 

For training and validation sets we synthesised $936,000$ and $104,000$ combinations of tissue properties, respectively. The sizes were chosen to obtain a $90/10$ split while ensuring a reasonable training time. NDI, ODI, and FWF were sampled uniformly in $[0.05,1]$, $[0.05,1]$, $[0, 0.95]$, as in previous studies \cite{Sam}.

For the test set, $500$ combinations of the scalar parameters were generated by gridding NDI, ODI, and FWF over the ranges $[0.05, 0.95]$, $[0.05, 0.95]$, and $[0.05, 0.45]$, respectively, using increments of $0.1$ \cite{Sam}. 

For each voxel we used 60 electrostatically optimised gradient directions per shell \cite{CaminoToolkit}. 
Once microstructural properties and gradient directions were synthesised, different choices for the b-values enabled the simulation of noise-free signals for three groups of train, validation, and test datasets. Each voxel in every set used two b-values, to match the hypernetwork input size. Specifically, train$_{\mathrm{1k2k}}$, validation$_{\mathrm{1k2k}}$ and test$_{\mathrm{1k2k}}$  used b-values $1000$ and $2000\;\mathrm{s/mm^2}$, train$_{\mathrm{1k3k}}$, validation\textsubscript{1k3k} and test$_{\mathrm{1k3k}}$ used b-values $1000$ and $3000\;\mathrm{s/mm^2}$, and each voxel in train$_{\mathrm{var}}$, validation$_{\mathrm{var}}$ and test$_{\mathrm{var}}$ had the lower and higher b-values sampled uniformly from $[700, 1200]$ and $[2000, 3000]$, respectively. The datasets with fixed b-value pairs were used to illustrate performance degradation on unseen b-values for an SCNN trained on a single protocol, and to assess how hSCNN mitigates this effect. The variable b-value dataset was used to evaluate generalisation in the standard ML sense, across protocols drawn from the training distribution.

Rician noise (SNR = 15) was added to all data, with 100 noise repetitions per voxel in the test sets. Finally, for each dataset, spherical harmonic coefficients were calculated per shell and concatenated as a matrix of size $45 \times 2$ (number of SH coefficients $\times$ number of shells), the input required by the SCNN architecture. 

\subsubsection{In-vivo Data}
In-vivo data was used only as test data and comprised masked brain volumes from two Human Connectome Project subjects. The acquisition included three shells ($b = 1000, 2000, 3000$ $\mathrm{s/mm^2}$), each with 90 gradient directions, and 18 $b=0$ images.

\subsection{Training Details} SCNN\textsubscript{1k2k}, SCNN$_{\mathrm{1k3k}}$, SCNN$_{\mathrm{var}}$, and hSCNN$_{\mathrm{var}}$ denote the model implementations trained on the datasets with the corresponding subscripts. The first three models were trained using Adam optimiser with a learning rate of $5\cdot10^{-4}$. For hSCNN$_{\mathrm{var}}$, we used a learning rate of $1\cdot10^{-4}$ and weight decay of $1\cdot10^{-6}$. All models were trained for 100 epochs with a batch size of 1000 using mean squared error (MSE) loss. Training required approximately 22 hours on a single NVIDIA GeForce RTX 4070 Ti SUPER GPU (CUDA 12.6).

\subsection{Evaluation Procedures}
\subsubsection{Evaluation on Synthetic Data}
\begin{figure}
    \centering
    \includegraphics[width=1\linewidth]{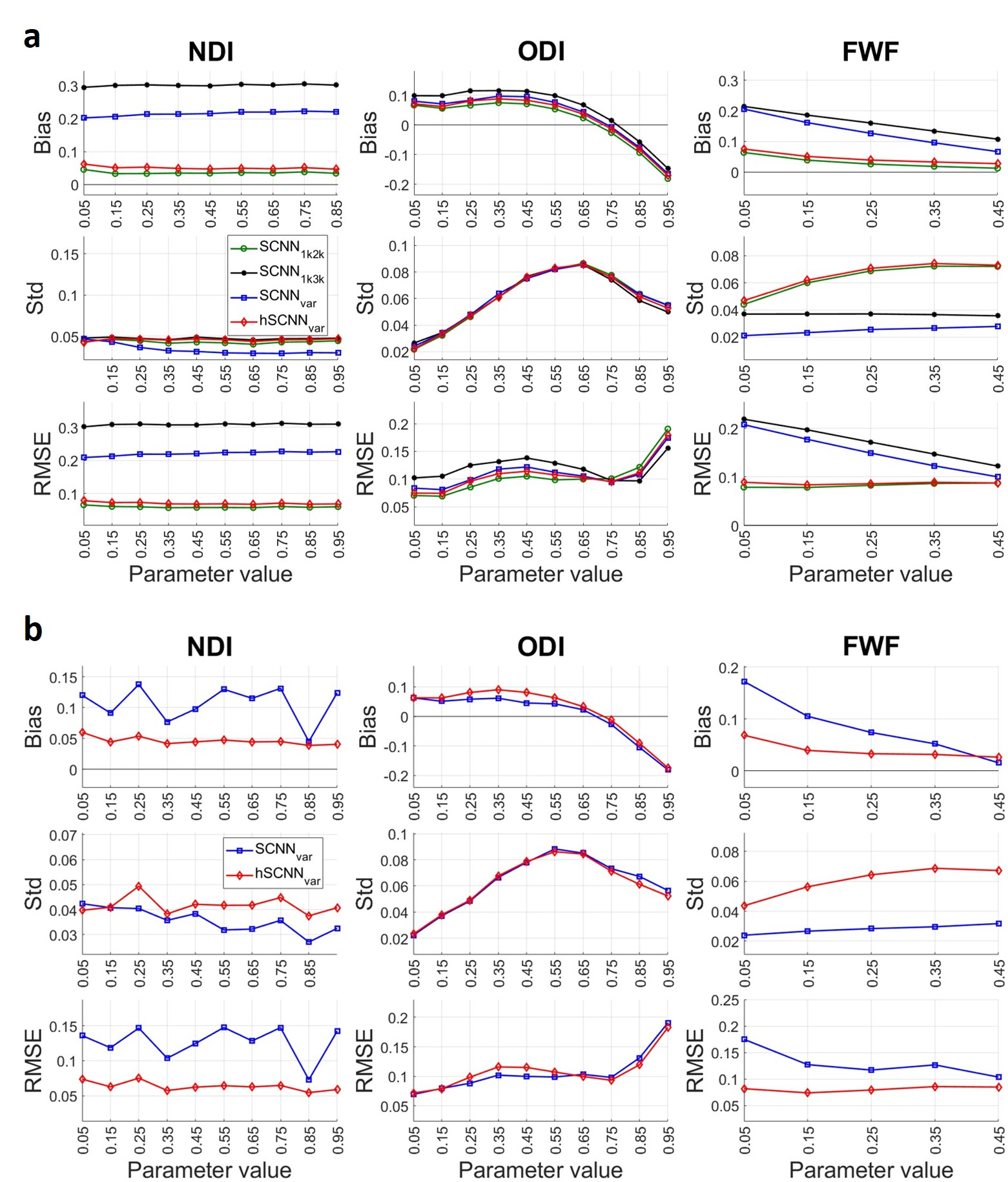}
    \caption{(a) Bias, standard deviation, and RMSE for the deep learning models on test$_{\mathrm{1k2k}}$. For each property, scores are marginalised over the remaining two properties. SCNN$_{\mathrm{1k2k}}$ and hSCNN\textsubscript{var} exhibit very similar performance, with lower bias and RMSE than both SCNN\textsubscript{var} and SCNN$_{\mathrm{1k3k}}$.
(b) Bias, standard deviation, and RMSE on test$_{\mathrm{var}}$. The hSCNN achieves lower bias and RMSE but higher standard deviation.}
    \label{fig:linePlots}
\end{figure}

All models were tested on test$_{\mathrm{1k2k}}$ and test$_{\mathrm{1k3k}}$ to assess their performances on specific protocols. SCNN$_{\mathrm{var}}$ and hSCNN\textsubscript{var} were additionally compared on test$_{\mathrm{var}}$. 
Inference with each deep learning model took $30$ seconds. Performances were compared via line plots of bias, standard deviation, and root mean squared error (RMSE) computed across noise trials (Figure \ref{fig:linePlots}). RMSE is indeed the most common metric for evaluating performance in parameter estimation \cite{Trainingdatadistribution}, but bias and standard deviation are useful complementary metrics \cite{Trainingdatadistribution,ML4dMRI}.

\subsubsection{Evaluation on In-Vivo Data}
Since the in-vivo data comprised three shells, while the deep learning models were trained on two shells, evaluation was performed on two subsets: one including only $b = 1000$ and $b = 2000\;\mathrm{s/mm^2}$, and the other with $b = 1000$ and $b = 3000\;\mathrm{s/mm^2}$. Inference times were 7.17 minutes (subject $100206$) and 5.67 minutes (subject $100307$) per subset.  A pseudo-ground truth was obtained performing conventional NODDI fitting, which used all available shells to ensure the highest quality reference. The fitting required 53.88 hours for subject $100206$ and 42.64 hours for subject $100307$. Qualitative assessment compared the parametric maps produced by the ML models with the NODDI estimates (Figures \ref{fig:QualReal_1k2k} and \ref{fig:QualReal_1k3k}). For each model and parameter, we also computed the mean RMSE and its standard deviation across datasets (Tables \ref{tab:QuantReal_1k2k} and \ref{tab:QuantReal_1k3k}) with respect to NODDI.

\section{Results}\label{Sec:Results}

\subsection{Results on Synthetic Data}
In all results, the differences introduced by the hypernetwork are most evident in NDI and FWF. These tissue properties are indeed more sensitive to signal scaling than ODI.

In particular, on test\textsubscript{1k2k} (Figure \ref{fig:linePlots}a), the metrics for hSCNN\textsubscript{var} closely reproduce those for SCNN\textsubscript{1k2k}, the model trained on the specific protocol of test\textsubscript{1k2k}. These two models achieve the lowest bias and RMSE for NDI and FWF. By contrast, the same metrics for SCNN\textsubscript{1k3k} and SCNN\textsubscript{var} are consistently two to three times larger, although these models yield a lower standard deviation on FWF. 

The results are similar on test\textsubscript{var} (Figure \ref{fig:linePlots}b), where SCNN\textsubscript{var} again shows higher bias and RMSE than hSCNN\textsubscript{var} across multiple ranges of NDI and FWF.

\subsection{Results on In-Vivo Data}
\begin{figure}[!ht]    
\centering
\includegraphics[height=0.8\textheight]{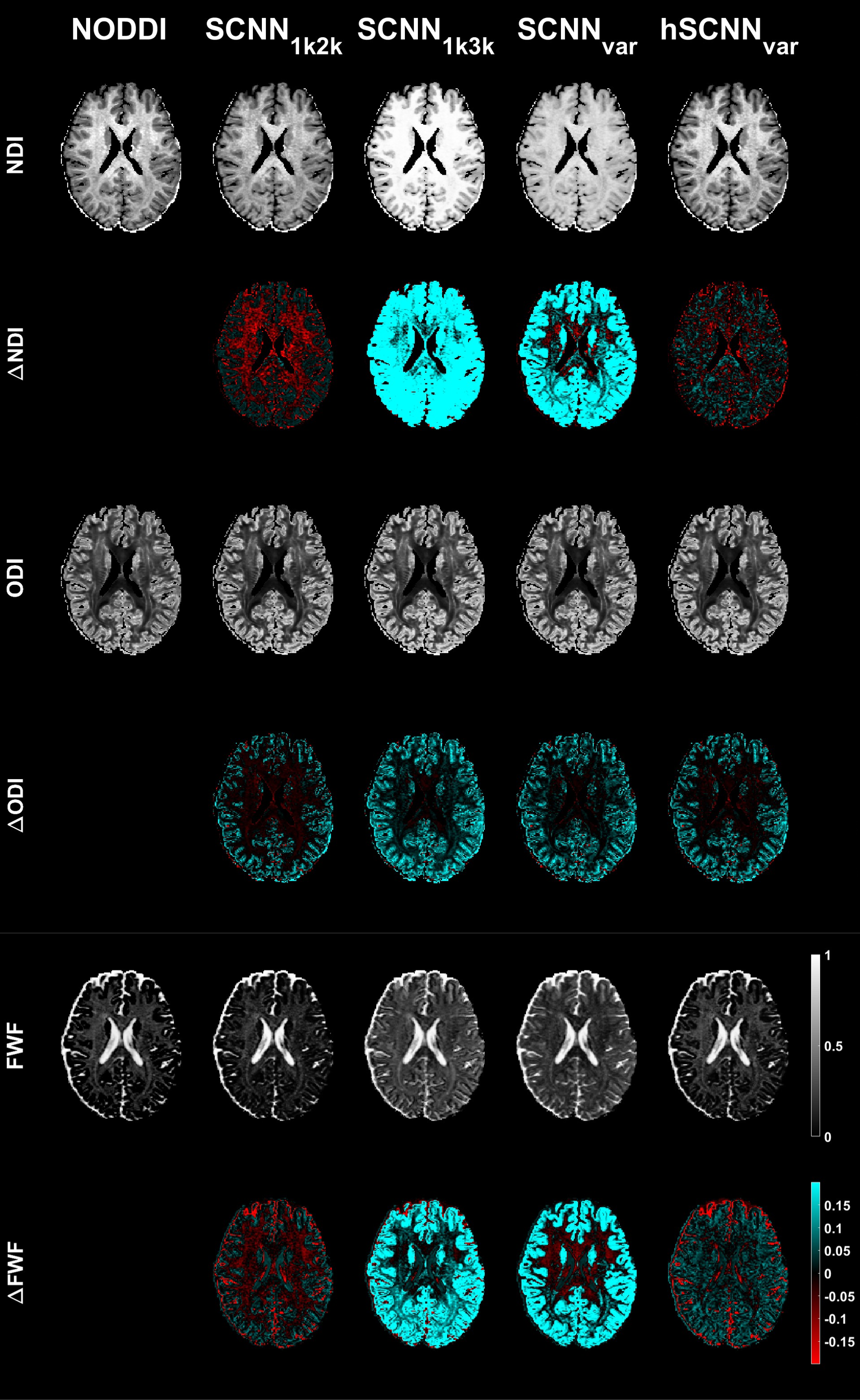}
    \caption{Qualitative results on the 72nd axial slice of the data subset from subject $100206$ using $b=1000$ and $b=2000\;\mathrm{s/mm^2}$. The first row shows model predictions for NDI; the second shows differences from conventional NODDI fitting. The same pattern is repeated for ODI and FWF in subsequent rows. SCNN\textsubscript{1k3k} shows evident overestimation. Although alleviated, high errors remain also for SCNN\textsubscript{var}. Instead, hSCNN$_{\mathrm{var}}$ consistently achieves high similarity to the NODDI predictions.} 
    \label{fig:QualReal_1k2k}
\end{figure}

On the real data subset with $b=1000$ and $b=2000\;\mathrm{s/mm^2}$, the qualitative evaluation (Figure \ref{fig:QualReal_1k2k}) shows that hSCNN\textsubscript{var} matches again the model trained specifically for the protocol of this test subset. Indeed, the maps returned by hSCNN\textsubscript{var} closely agree with SCNN\textsubscript{1k2k}. In addition, hSCNN\textsubscript{var} and SCNN\textsubscript{1k2k} display the largest similarity to NODDI. Instead, the maps returned by SCNN\textsubscript{1k3k} and SCNN\textsubscript{var} exhibit large areas of high overestimation, especially for NDI and FWF. 
These observations are further corroborated by the quantitative evaluation (Table \ref{tab:QuantReal_1k2k}). The table reports that hSCNN\textsubscript{var} and SCNN\textsubscript{1k2k} achieve the lowest mean RMSEs for NDI and FWF. Specifically, the mean RMSEs for hSCNN\textsubscript{var} are $0.0540$ on ODI and $0.0581$ on FWF, whereas those for SCNN\textsubscript{1k2k} are $0.0607$ on ODI and $0.0520$ on FWF. These values are all four to five times smaller than those for SCNN\textsubscript{1k3k}, whose mean RMSEs on NDI and FWF are $0.3063$ and $0.1911$, respectively.

\begin{figure}[!ht]    
\centering
\includegraphics[height=0.8\textheight]{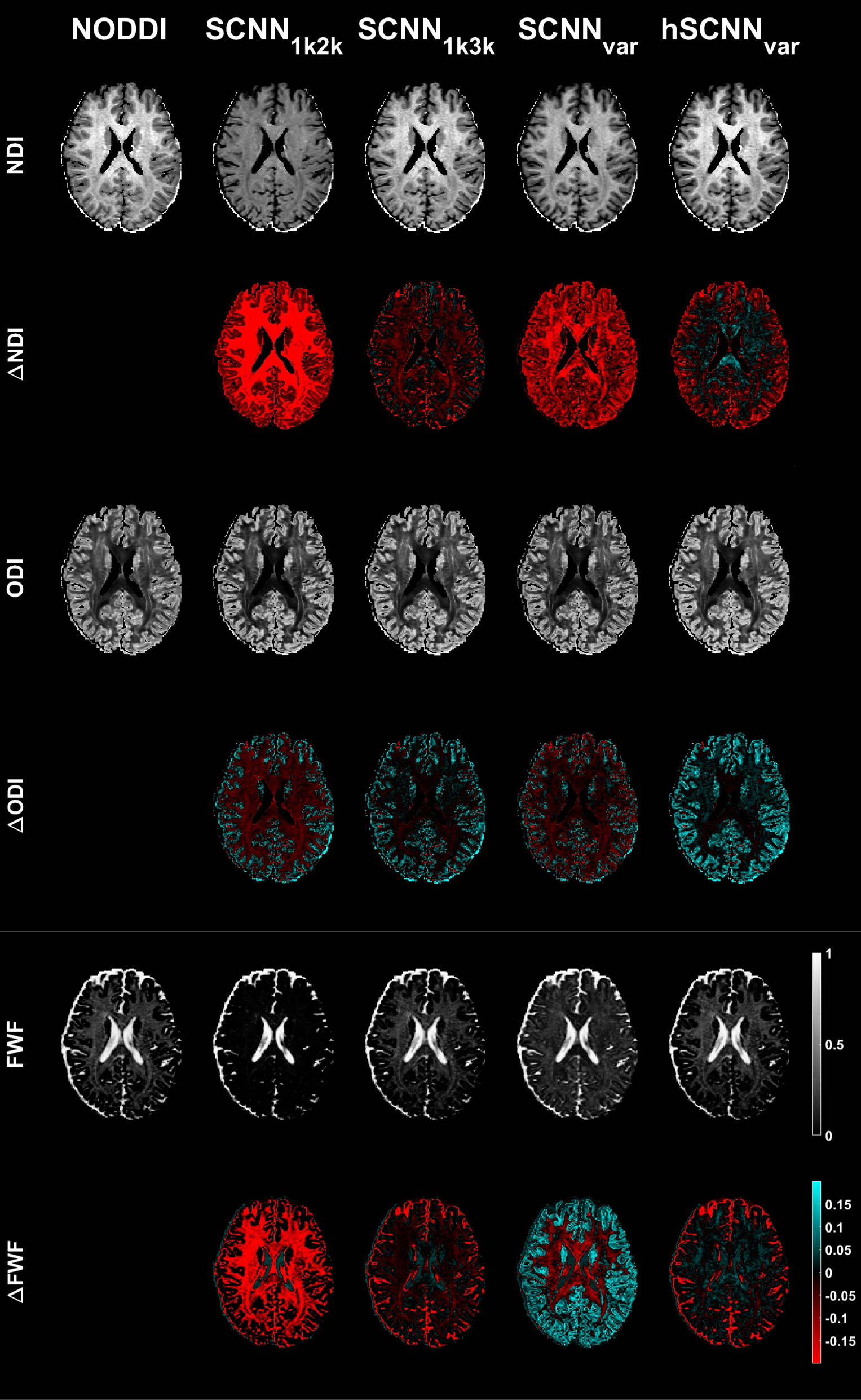}
    \caption{Qualitative results on the 72nd axial slice of the data subset from subject $100206$ using $b=1000$ and $b=3000\;\mathrm{s/mm^2}$. Mirroring Figure \ref{fig:QualReal_1k2k}, here hSCNN\textsubscript{var} closely agrees with SCNN\textsubscript{1k3k} and NODDI fitting, while SCNN\textsubscript{1k2k} displays clear underestimation of NDI and FWF.} 
    \label{fig:QualReal_1k3k}
\end{figure}

For $b=1000$ and $b=3000\text{ s/mm}^2$, the results follow a mirrored trend to the previous findings. Indeed, hSCNN\textsubscript{var} now matches SCNN\textsubscript{1k3k}, and both models in turn closely agree with NODDI. Conversely, high errors are observed for SCNN\textsubscript{1k2k} and SCNN\textsubscript{var} (Figure \ref{fig:QualReal_1k3k}), often as a consequence of conspicuous underestimations. This is again reflected in the quantitative evaluation (Table \ref{tab:QuantReal_1k3k}). Indeed, the lowest mean RMSEs for NDI and FWF are obtained by hSCNN\textsubscript{var} ($0.0764$ and $0.0736$, respectively) and SCNN\textsubscript{1k3k} ($0.0696$ and $0.0727$, respectively). In comparison, the mean RMSEs for SCNN\textsubscript{1k2k} reach $0.1899$ on NDI and $0.1315$ on FWF.

\section{Discussion and Conclusions}\label{Sec:Conclusion}
Across protocols, hSCNN\textsubscript{var} consistently matches the performance of an SCNN trained specifically for that protocol. Notably, for each test set, the errors of 
\begin{table}
\caption{RMSE (mean $\pm$ std) for each model and parameter across real data subsets with $b = 1000$ and $b = 2000\;\mathrm{s/mm^2}$. Bold font highlights the smallest mean RMSE for each parameter.}\label{tab:QuantReal_1k2k}
\centering
\setlength{\tabcolsep}{12pt} 
\begin{tabular}{|l|c|c|c|}
\hline
Model & NDI & ODI & FWF\\
\hline
SCNN\textsubscript{1k2k} & 0.0607 $\pm$ 0.0062 & \textbf{0.0626} $\pm$ 0.0003 & \textbf{0.0520} $\pm$ 0.0012\\
SCNN\textsubscript{1k3k} & 0.3063 $\pm$ 0.0107 & 0.0909 $\pm$ 0.0014 & 0.1911 $\pm$ 0.0107\\
SCNN\textsubscript{var}                  & 0.2034 $\pm$ 0.0082 & 0.0724 $\pm$ 0.0001 & 0.1757 $\pm$ 0.0096\\
hSCNN\textsubscript{var}                 & \textbf{0.0540} $\pm$ 0.0048 & 0.0680 $\pm$ 0.0002 & 0.0581 $\pm$ 0.0006\\
\hline
\end{tabular}
\end{table}
\begin{table}
\caption{RMSE (mean $\pm$ std) for each model and parameter across real data subsets with $b = 1000$ and $b = 3000\;\mathrm{s/mm^2}$. Bold font highlights the smallest mean RMSE for each parameter.}\label{tab:QuantReal_1k3k}
\centering
\setlength{\tabcolsep}{12pt} 
\begin{tabular}{|l|c|c|c|}
\hline
Model & NDI & ODI & FWF\\
\hline
SCNN\textsubscript{1k2k} & 0.1899 $\pm$ 0.0254 & 0.0685 $\pm$ 0.0003 & 0.1315 $\pm$ 0.0138\\
SCNN\textsubscript{1k3k} & \textbf{0.0696} $\pm$ 0.0048 & 0.0701 $\pm$ 0.0011 & \textbf{0.0727} $\pm$ 0.0036\\
SCNN\textsubscript{var}                  & 0.1150 $\pm$ 0.0103 & \textbf{0.0657} $\pm$ 0.0018 & 0.0887 $\pm$ 0.0000\\
hSCNN\textsubscript{var}                 & 0.0764 $\pm$ 0.0038 & 0.0827 $\pm$ 0.0012 & 0.0736 $\pm$ 0.0036\\
\hline
\end{tabular}
\end{table}
SCNN\textsubscript{var} and of the SCNN trained on the b-values of the other test set always appear visibly larger than the variations that would arise between NODDI fits on two separate protocols. This explicitly demonstrates that the baseline SCNNs do not generalise to unseen b-value pairs. In particular, the high errors displayed by SCNN\textsubscript{var} suggest that even training on many b-value pairs does not suffice to learn generalisable patterns. This is possibly a consequence of the degeneracy~\cite{Minore} of parameter estimation, for which two different b-values can produce similar signals despite large differences in the underlying tissue properties. 
Introducing explicit b-value dependence via the hypernetwork appears to resolve this degeneracy and attain protocol generalisation.

In conclusion, we developed a machine learning approach that simultaneously achieves rotational equivariance and generalisation across b-values and b-vectors. The proposed hSCNN is thus equipped with key properties for clinical dMRI parameter estimation and opens the door to ML applications in qMRI across diverse acquisitions. 

One limitation of this study is that using distinct parallel diffusivities for gray and white matter during data simulation and NODDI fitting could further improve the results. 

Future work could also investigate the role of parameter degeneracy in the performance differences between SCNN\textsubscript{var} and hSCNN\textsubscript{var}, extend the generalisation to variable numbers of b-values, and develop a self-supervised version of hSCNN\textsubscript{var} that removes the need for training labels.

\begin{credits}
\subsubsection{\ackname} Data were provided (in part) by the Human Connectome Project, WU-Minn Consortium (Principal Investigators: David Van Essen and Kamil Ugurbil; 1U54MH091657) funded by the 16 NIH Institutes and Centers that support the NIH Blueprint for Neuroscience Research; and by the McDonnell Center for Systems Neuroscience at Washington University.

\subsubsection{\discintname}
The authors have no competing interests to declare that are relevant to the content of this article.
\end{credits}

%
%
%
\bibliographystyle{splncs04}
\bibliography{mybibliography}

\end{document}